\documentstyle[prb,aps,multicol]{revtex}
  \tighten
  \begin{document}
  \draft
  \title{Wigner crystal model of counterion induced
    bundle formation of rod-like polyelectrolytes}
  \author {B. I. Shklovskii}
  \address{Theoretical Physics Institute, University of Minnesota,
  116 Church St. S.E., Minneapolis, Minnesota 55455}
  \maketitle
  \begin{abstract}
  A simple electrostatic theory of condensation of rod-like
  polyelectrolytes under influence of polyvalent ions is
  proposed. It is based on the idea that Manning condensation of ions
  results in formation of the Wigner crystal on a background of
  a bundle of rods. It is shown that, depending on a single
  dimensionless parameter, this
  can be the densely packed three-dimensional Wigner crystal
  or the two-dimensional crystal
  on the rod surfaces. For DNA the location of charge
  on the spiral results in a model of the one-dimensional Wigner crystal.
  It is  also argued that the Wigner crystal idea
  can be applied to self-assembly of other
  polyelectrolytes, for example, colloids and DNA-lipid complexes.
  \end{abstract}

  \begin{multicols}{2}

  Many rod-like polyelectrolytes, such as double helix DNA~\cite{Bloom},
  F-actin, microtubules, and tobacco mosaic virus~\cite{Janmey}
  are known to self-assemble into bundles of parallel
  densely packed rods. All these macromolecules are
  negatively charged and their lateral association is induced by
  $Z$-valent cations where $Z\geq 2$. For DNA this phenomenon is called
  condensation and is studied in connection with the dense packing
  of viral DNA.
According to the mean field Poisson-Boltzmann theory,
 two parallel rods should always repel each other.
Two physical phenomena which are not included in this theory
were suggested as possible reasons for the
puzzling attraction
~\cite{Bruinsma,Ray,Barrat,And,Manning,Oosawa,Rouzina,Parsegian,Andr}.
When distance between rods is large their attraction is related to the
  correlation of thermal fluctuations
of screening atmospheres of two rods.
 At smaller distances one should take into account that
 the Manning condensation of
ions on the surface of rods leads to strong
spatial correlations of ions or even to 
their crystallization.
In this situation, two crystals with proper phases attract each other.
Below I calculate the
binding energy of a dense bundle and therefore
talk only about the second mechanism. I also resolve
an important contradiction in the literature.
Most of the publications deal with two rods and calculate a pairwise
force acting between
them~\cite{Bruinsma,Ray,Barrat,And,Manning,Oosawa,Rouzina}.
  On the other hand, two recent publications~\cite{Parsegian,Andr}
  claim that bundles
  are formed by non-pairwise-additive interaction.

  The first goal of this paper
  is to present a theory of attractive interactions
  of rod-like polyelectrolytes, 
based solely on electrostatic interactions.
 It considers formation of densely packed
bundles in a very
  dilute solution of cylindrical
  rod-like molecules of the radius $r$
  and length $L\gg r$. It assumes that the cylindrical
  surface of a rod is negatively charged
  with the charge density
  $-e/b$ per unit length of the rod. Point-like positive ions with the
  charge $Ze$ are added to the solution.
 
The main idea of this paper is that a bundle of parallel rods
  can be considered as an uniform
 negative background at which condensed ions
  form the Wigner crystal. The cohesive energy
  of this crystal is the reason for the attraction and the
  bundle formation. (This idea is similar to
the theory of matter in a superstrong magnetic field
of a neutron star where electrons
 behave as negative rods and nuclei form the Wigner
 crystal on the background of electron bundles~\cite{Ruderman}.)
For the case in which
  ions can not penetrate the rods, the binding
  energy of the bundle per rod is calculated below
   as the function of the dimensionless parameter
  $Zb/r$. The result is that for $Zb/r \gg 1$, the
  three-dimensional densely-packed Wigner crystal
  is formed and interaction is
  not-pairwise-additive. In the opposite case, $Zb/r \ll 1$ the
  two-dimesional Wigner crystal appears on the surface of each rod
  and the interaction becomes short-range pairwise.
  My  results in the first case qualitatively resemble
  Ref.~\onlinecite{Andr},
  while in the second case this paper is
  close to Ref.~\onlinecite{Rouzina}, where idea of the surface
  Wigner crystal was originally suggested.
  None of these papers, however, produced simple analytical
  dependencies similar to those which I derive below
  using $Zb/r$ as a large or small parameter of the theory.

  The second half of this paper concentrates on the specifics of
  DNA where surface charges
  form a spiral. I show that this leads to the appearence of
   the one-dimesional spiral-like Wigner
crystal of ions on the surface of
  each rod. Interaction
  of such crystals determines the DNA condensation.
  It is also argued that the idea of a Wigner crystal
  can be applied not only to
  the rods, but also to the self-assembly of many other different
  polyelectrolytes. I apply this theory to colloids
  and complexes of DNA with cationic lipid membranes.

  Returning to the main problem of rods and point-like counterions,
  I argue that for sufficiently large dimensionless parameter $Z\xi$,
  where $\xi = e^2/ b \kappa k_B T$ and $\kappa$ is  the effective
  dielectric constant
  of the water-polymer system, rods condense in the maximum density
  cylindrical
  bundles of $N$ parallel molecules. At large  $Z\xi$ the charge
  of the bundle is almost completely compensated
  by the opposite charge of $NL/b$ of positive
  ions which experience the Manning condensation inside the bundle.
  To find a configuration of condensed ions,
  one can view the bundle as the uniformly charged
  cylindrical background. The potential energy of the Coulomb repulsion of
  ions in these conditions is much larger than their kinetic energy.
  Therefore, they form the Wigner crystal on the negative background.
  Assume first that $N$ is so large that the radius of the bundle
  $ rN^{1/2}$ is much larger than the lattice constant of the Wigner
  crystal $R$.
  In this case ions form a densely packed three-dimensional crystal.
  One can easily calculate  $R$ from the condition that the charge
  of the Wigner-Seitz cell, 
  $R^3/br^2$, equals $Z$. This gives
  \begin{equation}
  R \sim (Zbr^{2})^{1/3}.
  \label{lattice}
  \end{equation}
  Comparing $R$ with the bundle radius $rN^{1/2}$
  one finds that the three-dimensional case takes place
  at $N\gg N_c= (Zb/r)^{2/3}$. (In the opposite case
  $N\ll N_c$ ions form one-dimensional crystal as shown below).
  I assume that ions can not penetrate rods.
  Then ions can fit in the free space between rods without strong
  deformation of the three-dimensional crystal only at $R \gg  r$.
  According to
  Eq.~(\ref{lattice}) this inequality means that $Zb/r\gg 1$.
  Now we see how the important parameter $Zb/r$ appears in the theory.
  Assuming, for the beginning, that rods are narrow and
  weakly charged so that  $Zb/r\gg 1$ one finds that
  $ N_c \gg 1$ so that both situations $N\gg N_c$
  and $N\ll N_c$ have to be considered.

  The binding energy of a crystal per ion,
  $\varepsilon_{i}$,
  can be estimated as the energy of
  interaction of an ion with its Wigner-Seitz cell
  \begin{equation}
  \varepsilon_{i} \sim {{Z^{2}e^{2}}\over {\kappa R}}.
  \label{energy}
  \end{equation}
  Substituting $R$ from Eq.~(\ref{lattice}) one obtains
  \begin{equation}
  \varepsilon_{i} \sim {{Ze^{2}}\over{\kappa b}} (Zb/r)^{2/3}~~~~~~~~~~
  (Zb/r\gg 1;~~  N\gg N_c).
  \label{energy1}
  \end{equation}
  If $N\ll N_c$, the bundle is narrow and the ions 
form a one-dimensional Wigner crystal,
  lattice constant of which, $R$, is equal to $Zb/N$.
  Then, the energy of
  interaction of an ion with its Wigner-Seitz cell is
  \begin{equation}
  \varepsilon_{i} \sim {{Ze^{2}N} \over{\kappa b}}~ln(N_c/N)~~~~~~(Zb/r\gg
  1;~~  N\ll N_c).
  \label{energy2}
  \end{equation}
  It is easy to verify that  Eq.~(\ref{energy})
  matches  Eq.~(\ref{energy1}) at $N \sim N_c$.
  Thus, at $Zb/r\gg 1$,  the energy per ion, $\varepsilon_{i}$,
  grows linearly at small $N$ and reaches saturation at $N\gg N_c$.
  This is a clear demonstration of a
  non-pairwise additive interaction.
  Of course, the surface correction to the energy of the bundle provides
  some growth even  at $N \gg N_c$ and leads to formation
  of macroscopic bundles.

  In order to obtain the binding energy per molecule,
  Eq.~(\ref{energy2}) and   Eq.~(\ref{energy1}) have
  to be multiplied by the number of ions per molecule, $M= L/bZ$.
  This transition does not change the dependence on $N$.
  Therefore below I will continue to present
  results in the form of $\varepsilon_{i}$.

  Before switching to the more complicated case $Zb/r\ll 1$, I
  discuss the condition
  on the temperature $T$ or, in other words, on the parameter
   $Z\xi$ at which the
  suggested above theory is valid. Consider a large bundle
  of three-dimensional densely packed Wigner crystal, which stays neutral
  even at quite a small concentration of ions.
  If  $k_B T \ll \varepsilon_i$ the
  thermal motion  can be neglected  and my theory is valid. Using
  Eq.~(\ref{energy1})
  one can  rewrite inequality  $k_B T \ll \varepsilon_i$ as
  \begin{equation}
  Z\xi \gg (r/Zb)^{2/3}.
  \label{xi}
  \end{equation}
  At $Zb/r \gg 1$ this inequality is much weaker than the standard
  condition
  $Z\xi \gg 1$ of
  the Manning condensation of ions with charge $Z$
   at an isolated molecule
   (it follows from Eq.~(\ref{energy2}) at $N=1$),
   because of the simultaneous interaction of each ion with many rods.
   Even if
   one uses  $\kappa =81$ of pure water, $\xi \sim 4$ at room temperature and
   the condition of Eq.~(\ref{xi}) is easily fulfilled.

  It is obvious that wnen Eq.~(\ref{xi}) is valid, the maximum density bundle
  is more strongly bounded than one with the smaller density.
  Indeed, the decreasing density results in the increase of the lattice
  constant
  $R$ and, according to Eq.~(\ref{energy}),
   substantially diminishes $\varepsilon_i$.
  At the same time the increase of the entropy term
  in free energy can not compensate for this loss in the binding energy.

  As stated above, this theory works if $k_B T \ll \varepsilon_i$.
  Actually the Wigner crystal melts when $k_B T$
  is yet numerically much smaller than $\varepsilon_i$, so
   in some experimental conditions or
  for low dimensionalities (see below)
  one  deals with strongly correlated liquid, not a crystal. This, however,
  does not
  change the estimate for its correlation energy and for $\varepsilon_i$
  and leaves unchanged the validity criterion of the theory.
The fact that it is the short range order of ions
which is responsible for the attraction of rods at a finite temperature
was emphasized in  Ref.~\onlinecite{Bruinsma}.
Below, I continue to use the Wigner crystal
 language because it creates a simpler image.

   Consider now more strongly charged and thicker
 rods for which $Zb/r \ll 1$.
  In this case $N_c \ll 1$ and, therefore, one has to deal only with a
three-dimensional problem.
  However, the rods are so thick that their radius
  is larger than the lattice constant of
  the optimal densely packed crystal. If, as we assumed,
  ions can not penetrate rods, the optimal densely packed
  crystal can not be formed. Under this
  restriction, ions condense at the surface of each rod forming
  the two-dimensional Wigner crystal~\cite{Rouzina} with the lattice
  constant
  \begin{equation}
  R \sim (Zbr)^{1/2}.
  \label{R2}
  \end{equation}
  An insulated rod has a similar crystal at its surface.
  The binding energy of the bundle originates in narrow contact stripes
  where rods pairwise contact each other so that their Wigner crystals
   overlap.  The width $W$ of this stripe  will be calculated below.
  Inside the stripe,  densities of both negative  background and
  positive ions are doubled and
  the local lattice constant $R_s$ becomes smaller ($R_s=R/\sqrt{2}$).
  As a result, according to Eq.~(\ref{energy}), the cohesive energy
of the crystal per ion of the stripe becomes larger
  than for the case of the two separated rods. Combining
  Eq.~(\ref{energy})
  and Eq.~(\ref{R2}) one finds that the binding energy of
the bundle  per one ion of the stripe
 equals $Z^{3/2}e^2/(br)^{1/2}$.  To get $\varepsilon_i$ one must multiply
  this energy by the fraction, $p \sim W/r$, of all ions of the bundle which
  reside in the contact stripes.
   The  width $W$ can be found from the condition
  that at the distance $W$, from the line of contact
  the surfaces of two contacting rods
  diverge from each other at the distance of the order of
  $R$. Indeed, the interaction between two-dimensional Wigner
  crystals is exponentially weak if the distance between
   parallel planes in which they are situated is larger
   than their lattice constant $R$.
  A simple geometrical estimate gives $W\sim (rR)^{1/2}$.
   Finishing the calculation of $\varepsilon_{i}$ one obtains
  \begin{equation}
  \varepsilon_{i} \sim {{Ze^{2}}\over{\kappa b}} (Zb/r)^{3/4}~~~~
  (Zb/r\ll 1;~~  N\gg 1).
  \label{energy3}
  \end{equation}
  Together Eq.~(\ref{energy3}) and  Eq.~(\ref{energy1})
  give binding energy per ion
  of a three-dimensional crystal at all values
   $Zb/r$. At $Zb/r=1$
  they obviously match each other. At $Zb/r \ll 1$
  energy given by Eq.~(\ref{energy3})
  is smaller than the one from
  Eq.~(\ref{energy1}). This is a natural result
  of the restriction that ions
  do not penetrate rods.

  Consider now the validity of the main assumption, that an array of
  negative discrete charges on
  the rod surface
  can be effectively replaced
  by an uniform negative background.
  This idea works
  exactly only in the limit $Z\gg 1$.
  On the other hand, it fails
  at $Z=1$ because  in low temperature limit
  all ions and discrete negative charges
 form neutral Bjerum pairs and
  nothing depends on the mutual positions of
  the rods.  Thus, $\varepsilon_{i}=0$ at $Z=1$.
  What happens at $Z \geq 2$
  depends on the spatial distribution of the discreet negative charges.
  If they are distributed randomly, then at $Z=2$
  Eqs.~(\ref{energy1}),~(\ref{energy2})
  and~(\ref{energy3}) overestimate $\varepsilon_{i}$ roughly 
by a factor of 2, while at $Z \geq 3$ they work well.
On the other hand, when negative
charges are clustered, $\varepsilon_{i}$
can be much lower even at $Z = 3$.
Imagine, for example, that negative charges form compact triplets. 
Each one would be neutralized by an ion and $\varepsilon_{i}$ would vanish.
More interesting effect of concentration of negative charges in DNA is discussed below.

  But first assume that the surface of
  DNA is uniformly negatively charged. For DNA $b=0.17$nm and $r=1$nm, so
  that
  $Z=6$ is the border between range of validity of
  Eq.~(\ref{energy3}) and Eq.~(\ref{energy1}).
  This means that for DNA in this model, attraction is due to the short range
  pairwise interaction of molecules induced by
  correlation between
  their surface Wigner crystals. This
  conclusion is in agreement with Ref.~\onlinecite{Rouzina}.

  Recall now that in DNA, negative charges are located along two
  spirals which are separated by wide and narrow grooves.
  For simplicity, assume that the width of the narrow one is zero
  what leaves only one spiral of the negative charge with the double
  density.
   (This approximation, of course, overestimates the inhomogeniety of
  the negative charge distribution and can be easily
   avoided with minor change in the result.) Ions
  tend to concentrate on the same spiral and form the one-dimensional Wigner
  crystal along it.
  When two parallel DNA rods touch each other by these
  crystaline spirals
  (for this purpose one rod should be shifted 
  along their direction by the
  half of the helical period) they create a spot with larger
 binding energy per ion. Thus, attraction is produced
in a way very similar to the case of contacting
  two-dimensional crystals discussed above. Extending this analogy
  for two contacting DNA molecules, one can introduce
  the contact stripe where attraction is generated.
  The only difference is that the width of this stripe $W$ is smaller for DNA,
  because the Wigner crystal spirals of the two contacting
   DNA molecules diverge not only
  in the plain perpendicular to the rods but also along direction of
  the rods (spirals cross at a finite angle). One can find
  that $W\sim Rr/nb \sim R$,
  where $R=\pi r/n$ is a period of the one-dimensional crystal and $n=10$
  is the number of base pairs in the helix period.
   Thus, roughly speaking, only one ion per helix period
  adjusts its position due to the contact
  and contributes to the binding energy of the two contacting
  rods. Multiplying  Eq.~(\ref{energy}) by the fraction
  of such ions $R/{2\pi r}$ one obtains
  \begin{equation}
  \varepsilon_{i} \sim {{Ze^{2}}\over{\kappa b}}{{Zb}\over{2\pi r}}.
  \label{energy4}
  \end{equation}
  At $Zb/r\ll 1$, Eq.~(\ref{energy4})  gives smaller energy than
  Eq.~(\ref{energy3}).
  This is an effect of concentration of negative charges on
  the spiral.
Note that this effect is exactly opposite to the prediction of
Ref.~\onlinecite{Leikin}.
The reason for this difference is that the authors of
Ref.~\onlinecite{Leikin} assumed that charge of ions is uniformly
distributed over DNA surface, so that negative and positive charges are
strongly separated. Such a separation of charge costs 
a large electrostatic energy and can happen 
only if nonelectrostatic forces  dominate. 
These forces are beyond the scope of my work.
 
One might wonder whether each contacting
  pair of DNA molecules in a bundle can
  gain the energy given by Eq.~(\ref{energy4}).
 This can be easily done in a square
  lattice of rods, with one square sublattice shifted
  along the rod direction by the half of the  helical period.
  In a densely packed hexagonal lattice one can
  shift in the same way every second layer, so that each rod
  is attracted to 4 its neighbours.
  Thus in the approximation of one spiral both
  lattices have the same energy.

  Note that literally speaking this theory is
  applicable only to molecules of DNA which are
  shorter than its persistent length. It is known that in a weak DNA
  solution of longer molecules, each molecule
  self-assembles into a toroidal particle, where locally
  strands of all turns are parallel to each other. Theory of this paper
  is applicable to toroidal
particles as well. 
  Even longer DNA molecules condense in more
  complex globules~\cite{Khokhlov}.
 If such a globula consists of condensed bundles 
this theory can still work.

  Although three-dimensional densely 
packed Wigner crystal of ions is not realized in DNA bundles
(for $Z <6$), many other rod-like polyelectrolytes are known.  Some
of them are narrower or more
weakly charged  so that $Zb/r \gg 1$ and
three-dimensional crystal of ions should appear.
PPP3 with $b/r \sim 2$
is a good example~\cite{Khokhlov1}.

  I would like to show now that the idea of Wigner
   crystal (or strongly correlated liquid) of ions
  for description of their Manning condensation can
  be applied not only to rods but to quite a variety
  of systems. Imagine, for example, a
  rigid plane polyelecrolyte
   in the presence of
   counterions with a large charge $Z$. Such a system
  self-assembles in a dense stack of parallel sheets
  which provides
  uniform background for the Wigner crystal of ions.
  Once more, depending on the thickness of the sheets
  and their surface charge density, one can
  talk about the three-dimensional
  Wigner crystal of ions or about two-dimensional crystals in
  gaps between sheets.

  Counterions  should not necessarily be small.
  The role of a counterion can be played
  by a strongly charged
  rodlike polymer while the uniform background
  is provided by the stack of positive parallel sheets.
Complexes of DNA with cationic lipid membranes
  are good examples of such a system~\cite{Safinya}.
  All the rods orient themselves parallel to each other and to the sheets.
  Their ordering in a plane
  perpendicular to their long axis
  depends on the sheet thickness. At a small enough sheet thickness, the circular
  cross-sections of rods are arranged  in
  the two-dimensional Wigner crystal similar to vortex lattice in superconductors,
  while at larger thicknesses they form parallel one dimensional
  crystals in the gaps between sheets.

  Returning to small counterions, consider
 the self-assembly of rigid spherical
  polyelectrolytes. Thinking about colloids,
  assume that the surface of a sphere is charged negatively
  and  positive ions can not penetrate it.
 If spheres are large enough,
  ions form the two-dimensional Wigner crystal
  at the surface
  of a sphere. This leads to contact attraction of spheres
  similar to the one calculated above for  two contacting
  cylinders. The binding energy of two spheres
  with the radius $r$ and the
  charge $-Qe$, originates in the contact disc
  (an analog of the contact stripe for two cylinders)
  of the radius $(rR)^{1/2}$, where $r$ is the sphere radius, $R$
  is the lattice constant of the Wigner crystal (see computer
  simulation of these discs in Ref.~\onlinecite{Colloid}).
  The additional energy  per ion of the disc is gained,
  because the disc has a denser
  background and a larger ion density than the rest of the
  surfaces of the spheres.
  Once more $\varepsilon_{i}$ is given by Eq.~(\ref{energy}).
  Multiplying
  this energy by the number of ions in such a
  disc, $r/R$, and substituting $R\sim r(Z/Q)^{1/2}$
  one gets a surprisingly simple expression for
  the binding energy of two spheres
  \begin{equation}
  \varepsilon \sim {{ZQe^{2}}\over{\kappa r}}~,
  \label{energy5}
  \end{equation}
  which looks like a Coulomb attraction energy of the spherical colloid
  particle and a single ion at its surface. 
 Eq.~(\ref{energy5})
fails with the decrease of the ion
concentration in the solution, $c$, when each
sphere acquires the net negative charge sufficient
to overcome calculated attraction. 
This happens at $k_BTln(1/c) \sim Z^{2}e^{2}/\kappa R$. 
At lower temperature Eq.~(\ref{energy5}) gives energy
which is larger than proportional to $T$ depletion energy of
Ref.~\onlinecite{Colloid}. 

  In conclusion, this paper studies the attraction and self-assembly
  of rigid polyelectrolytes  due to strong correlations in positions
  of polyvalent counterions. To describe these correlations,
  I adopt the notion of the Wigner crystal.
  For rods I find that,
  depending on the value of a single dimensionless
  combination, $Zb/r$,  optimal cofiguration of ions on
  the background of a bundle is the three-dimensional
  densely packed Wigner crystal or the two-dimensional
  crystal on the surface of rods.  When applying this theory
  to DNA condensation I introduce the notion of
  one-dimensional spiral Wigner crystal
  and estimate the binding energy of two DNA helixes
  due to the contact of these crystals. I also calculate the
  binding energy of two colloidal particles created by
  the contact of Wigner crystals of ions on their surfaces.
  Future developments of this work should include such factors as 
  behaviour
  of the dielectric constant of water at small distances,
  the finite size  of counterions, and more realistic
  distributions of polyelectrolyte charges.


  \acknowledgements

  I am grateful to V. A. Bloomfield, A. M. Dykhne, 
  M. D. Frank-Kamenetskii, A. Yu. Grosberg, R. Holyst,
  A. R. Khokhlov, and I. Rouzina
  for valuable discussions.
  This work was started in Aspen Center for Physics and was
  supported by NSF DMR-9616880.


  \end{multicols}
  \end{document}